\documentclass[conference]{IEEEtran}
\IEEEoverridecommandlockouts

\usepackage{amssymb}
\usepackage{amsmath}
\usepackage{flushend}
\usepackage{pgfplots}
\usepackage{multirow}
\usepackage{comment}
\usepackage{booktabs}
\usepackage{comment}
\usepackage{adjustbox}
\usepackage{bm}

\begin{document}
\pgfplotsset{ymin=0}

\title{Additive Margin SincNet for Speaker Recognition
}

\author{
\IEEEauthorblockN{Jo\~ao Ant\^onio Chagas Nunes}
\IEEEauthorblockA{\textit{Centro de Inform\'{a}tica}\\
\textit{Universidade Federal de Pernambuco}\\
50.740-560, Recife, PE, Brazil\\
jacn2@cin.ufpe.br}\\
\and
\IEEEauthorblockN{David Mac\^edo}
\IEEEauthorblockA{\textit{Centro de Inform\'{a}tica}\\
\textit{Universidade Federal de Pernambuco}\\
50.740-560, Recife, PE, Brazil\\
dlm@cin.ufpe.br}\\
\and
\IEEEauthorblockN{Cleber Zanchettin}
\IEEEauthorblockA{\textit{Centro de Inform\'{a}tica}\\
\textit{Universidade Federal de Pernambuco}\\
50.740-560, Recife, PE, Brazil\\
cz@cin.ufpe.br}\\
}


\maketitle


\begin{abstract}

Speaker Recognition is a challenging task with essential applications such as authentication, automation, and security. The SincNet is a new deep learning based model which has produced promising results to tackle the mentioned task. To train deep learning systems, the loss function is essential to the network performance. The Softmax loss function is a widely used function in deep learning methods, but it is not the best choice for all kind of problems. For distance-based problems, one new Softmax based loss function called Additive Margin Softmax (AM-Softmax) is proving to be a better choice than the traditional Softmax. The AM-Softmax introduces a margin of separation between the classes that forces the samples from the same class to be closer to each other and also maximizes the distance between classes. In this paper, we propose a new approach for speaker recognition systems called AM-SincNet, which is based on the SincNet but uses an improved AM-Softmax layer. The proposed method is evaluated in the TIMIT dataset and obtained an improvement of approximately 40\% in the Frame Error Rate compared to SincNet.

\end{abstract}


%
\IEEEpeerreviewmaketitle

\section{Introduction}
\label{introduction}

Speaker Recognition is an essential task with applications in biometric authentication, identification, and security among others \cite{Beigi}. The field is divided into two main subtasks: Speaker Identification and Speaker Verification. In Speaker Identification, given an audio sample, the model tries to identify to which one in a list of predetermined speakers the locution belongs. In the Speaker Verification, the model verifies if a sampled audio belongs to a given speaker or not. Most of the literature techniques to tackle this problem are based on $i$-vectors methods \cite{Dehak10frontend}, which extract features from the audio samples and classify the features using methods such as PLDA \cite{Prince2007ProbabilisticLD}, heavy-tailed PLDA \cite{htPLDA}, and  Gaussian PLDA \cite{GaussPLDA}.

Despite the advances in recent years \cite{oldmethods, oldmethods2, oldmethods5, oldmethods6, oldmethods7, advancesDeep, advancesDeep2}, Speaker Recognition is still a challenging problem. In the past years, Deep Neural Networks (DNN) has been taking place on pattern recognition tasks and signal processing. Convolutional Neural Networks (CNN) have already show that they are the actual best choice to image classification, detection or recognition tasks. In the same way, DNN models are being used combined with the traditional approaches or in end-to-end approaches for Speaker Recognition tasks \cite{end-to-end, end-to-end2, end-to-end3}. In hybrid approaches, it is common to use the DNN model to extract features from a raw audio sample and then encode it on embedding vectors with low-dimensionality which samples sharing common features with closer samples. Usually, the embedding vectors are classified using traditional approaches.

The difficult behind the Speaker Recognition tasks is that audio signals are complex to model in low and high-level features that are discriminant enough to distinguish different speakers. Methods that use handcrafted features can extract more human-readable features and have a more appealing approach because humans can see what the method is doing and which features are used to make the inference. Nevertheless, handcrafted features lack in power. In fact, while we know what patterns they are looking for, we have no guarantee that these patterns are the best for the job. On the other hand, approaches based on Deep Learning have the power to learn patterns that humans may not be able to understand, but usually get better results than traditional methods, despite having more computational cost to training. 

\begin{figure*}[t]
\centering
\includegraphics[width=0.8\textwidth]{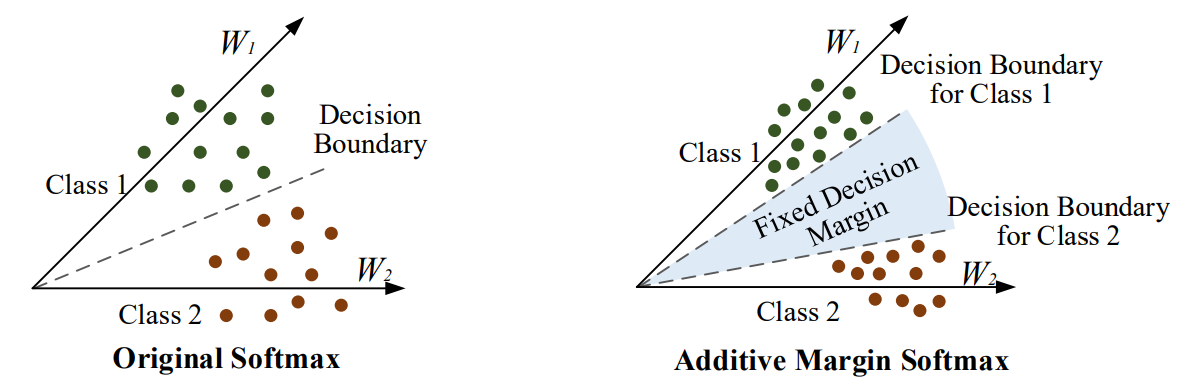}
\caption{Comparison between Softmax and AM-Softmax \cite{AM-Softmax}.}
\label{fig:SoftmaxVSAMSoftmax}
\end{figure*}

A promising approach to Speaker Recognition based on Deep Learning is the SincNet model \cite{SincNet} that unifies the power of Deep Learning with the interpretability of the handcrafted features. SincNet uses a Deep Learning model to process raw audio samples and learn powerful features. Therefore, it replaces the first layer of the DNN model, which is responsible for the convolution with parametrized sinc functions. The parametrized sinc functions implement band-pass filters and are used to convolve the waveform audio signal to extract basic low-level features to be later processed by the deeper layers of the network. The use of the sinc functions helps the network to learn more relevant features and also improves the convergence time of the model as the sinc functions have significantly fewer parameters than the first layer of traditional DNN. At the top of the model, the SincNet uses a Softmax layer which is responsible for mapping the final features processed by the network into a multi-dimensional space corresponding to the different classes or speakers.

The Softmax function is usually used as the last layer of DNN models. The function is used to delimit a linear surface that can be used as a decision boundary to separate samples from different classes. Although the Softmax function works well on optimizing a decision boundary that can be used to separate the classes, it is not appropriate to minimize the distance from samples of the same class. These characteristics may spoil the model efficiency on tasks like Speaker Verification that require to measure the distance between the samples to make a decision. To deal with this problem, new approaches such as Additive Margin Softmax \cite{AM-Softmax} (AM-Softmax) are being proposed. The AM-Softmax introduces an additive margin to the decision boundary which forces the samples to be closer to each other, maximizing the distance between the classes and at the same time minimizing the distance from samples of the same class.

In this paper, we propose a new method for Speaker Verification called Additive Margin SincNet (AM-SincNet) that is highly inspirited on the SincNet architecture and the AM-Softmax loss function. In order to validate our hypothesis, the proposed method is evaluated on the TIMIT \cite{timit} dataset based in the Frame Error Rate. The following sections are organized as: In Section \ref{related_work}, we present the related works, the proposed method is introduced at Section \ref{additive_margin_sincnet}, Section \ref{experiments} explains how we built our experiments, the results are discussed at Section \ref{results}, and finally at Section \ref{conclusion} we made our conclusions. 

\section{Related Work}
\label{related_work}

For some time, $i$-vectors \cite{Dehak10frontend} have been used as the state-of-the-art feature extraction method for speaker recognition tasks. Usually, the extracted features are classified using PLDA \cite{Prince2007ProbabilisticLD} or other similar techniques, such as heavy-tailed PLDA \cite{htPLDA} and Gauss-PLDA \cite{GaussPLDA}. The intuition behind these traditional methods and how they work can be better seem in \cite{TraditionalMethodsReview}. Although they have been giving us some reasonable results, it is clear that there is still room for improvements \cite{TraditionalMethodsReview}. 

Recently, neural networks and deep learning techniques have shown to be a particularly attractive choice when dealing with feature extraction and patterns recognition in the most variety of data \cite{Goodfellow-et-al-2016, RavanelliBOB17}. For instance, CNNs are proving to produce a high performance on image classification tasks. Moreover, deep learning architectures \cite{deep1, deeo2} and hybrid systems \cite{hybird1, hybrid2, hybrid3, hybrid4, hybrid5} are higher quality results on processing audio signals than traditional approaches. As an example, \cite{triplet} built a speaker verification framework based on the Inception-Resnet-v1 deep neural network architecture using the triplet loss function.

SincNet \cite{SincNet} is one of these innovative deep learning architecture for speaker recognition which uses parametrized sinc functions as a foundation to its first convolutional layer. Sinc functions are designed to process digital signals just like audio, and thus the use of them as the first convolutional layer helps to capture more meaningful features to the network. Additionally, the extracted features are also more human-readable than the ones obtained from ordinary convolutions.

Besides, the sinc functions reduce the number of parameters on the SincNet first layer because each sinc function of any size only have two parameters to learn against $L$ from the conventional convolutional filter, where $L$ is the size of the filter. As a result, the sinc functions enables the network to converge faster. Another advantage of the sinc functions is the fact that they are symmetric, which means that we can reduce the computational effort to process it on $50\%$ by simply calculating half of the filters and flipping it to the other side.

The first layer of SincNet is made by 80 filters of size 251, and then it has two more conventional convolutional layers of size five with 60 filters each. Normalization is also applied to the input samples and the convolutional layers, the traditional and the sinc one. After that, the result propagates to three more fully connected layers of size 2048, and it is normalized again. The hidden layers use the Leaky ReLU \cite{leaky-ReLU} as the activation function. The sinc convolutional layer is initialized using mel-scale cutoff frequencies. On the other hand, the traditional convolutional layers together with the fully connected layers are initialized using $Glorot$ scheme. Finally, a Softmax layer provides the set of posterior probabilities for the classification.

\section{Additive Margin SincNet}
\label{additive_margin_sincnet}

The AM-SincNet is built by replacing the softmax layer of the SincNet with the Additive Margin Softmax \cite{AM-Softmax}. The Additive Margin Softmax (AM-Softmax) is a loss function derived from the original Softmax which introduces an additive margin to its decision boundary.

The additive margin works as a better class separator than the traditional decision boundary from Softmax. Furthermore, it also forces the samples from the same class to become closer to each other thus improving results for tasks such as classification and verification. The AM-Softmax equation is written as:

\begin{equation}
\label{eq:AM-Softmax}
Loss = -\frac{1}{n} \sum^n_{i=1}log \frac{\phi_i}{\phi_i + \sum^c_{j=1, j \neq y_i }exp(s(W_{j}^T f_i))}
\end{equation}
\begin{equation}
\phi_i = exp(s(W_{y_i}^T f_i-m))
\end{equation}

In the above equation, W is the weight matrix, and $f_i$ is the input from the $i$-th sample for the last fully connected layer. The $W_{y_i}^T f_i$ is also known as the target logit for the $i$-th sample. The $s$ and $m$ are the parameters responsible for scaling and additive margin, respectively. Although the network can learn $s$ during the optimization process, this can make the convergence to be very slow. Thus, a smart choice is to follow \cite{AM-Softmax} and set $s$ to be a fixed value. On the other hand, the $m$ parameter is fundamental and has to be chosen carefully. On our context, we assume that both $W$ and $f$ are normalized to one. Figure \ref{fig:SoftmaxVSAMSoftmax} shows a comparison between the traditional Softmax and the AM-Softmax.

\begin{figure}[ht]
\centering
\includegraphics[width=\columnwidth]{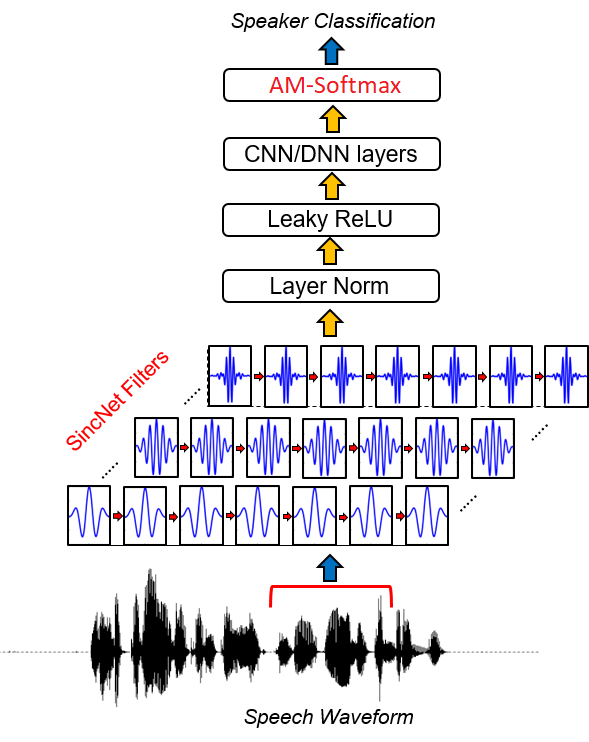}
\caption{Illustration of the proposed AM-SincNet architecture. Adapted from the original SincNet \cite{SincNet}}
\label{fig:AM-SincNet}
\end{figure}

The SincNet approach has shown high-grade results on the speaker recognition task. Indeed, its architecture has been compared against ordinary CNNs and several other well-known methods for speaker recognition and verification such as MFCC and FBANK, and, in every scenario, the SincNet has overcome alternative approaches. The SincNet most significant contribution was the usage of sinc functions as its first convolutional layer. Nevertheless, to calculate the posterior probabilities over the target speaker, SincNet applies the Softmax loss function which, despite being a reasonable choice, is not particularly capable of producing a sharp distinction among the class in the final layer. Thus, we have decided to replace the last layer of SincNet from Softmax to AM-Softmax. Figure \ref{fig:AM-SincNet} is a minor modification of the original SincNet image that can be found in \cite{SincNet} which shows the archtecture of the proposed AM-SincNet.

\section{Experiments}
\label{experiments}

The proposed method AM-SincNet has been evaluated on the well known TIMIT dataset \cite{timit}, which contains audio samples from 630 different speakers of the eight main American dialects and where each speaker reads a few phonetically rich sentences. We used the same pre-processing procedures as \cite{SincNet}. For example, the non-speech interval from the beginning and the end of the sentences were removed. Following the same protocol of \cite{SincNet}, we have used five utterances of each speaker for training the network and the remaining three for evaluation. Moreover, we also split the waveform of each audio sample into 200ms chunks with 10ms overlap, and then these chunks were used to feed the network.

For training, we configured the network to use the RMSprop as optimizer with mini-batches of size 128 along with a learning rate of $lr\!=\!0.001$, $\alpha\!=\!0.95$, and $\epsilon\!=\!10^{-7}$. The AM-Softmax comes with two more parameters than the traditional Softmax, and the new parameters are the scaling factor $s$  and the margin size $m$. As mentioned before, we set the scaling factor $s$ to a fixed value of 30 in order to speed up the network training. On the other hand, for the margin parameter $m$ we carefully did several experiments to evaluate the influence of it on the Frame Error Rate (FER).

We also have added an $epsilon$ constant of value $10^{-11}$ to the AM-Softmax equation in order to avoid a division by zero on the required places.
For each one of the experiments, we trained the models for exactly 352 epochs as it appeared enough to exploit adequately the different training speed presented by both competing models. To run the experiments, we used an NVIDIA Titan XP GPU, and the training process lasts for about four days. The experiments performed by this paper may be reproduced by using the code that we made available online at the GitHub\footnote{https://github.com/joaoantoniocn/AM-SincNet}.

\begin{table*}[ht]
\centering
\caption{SincNet and AM-SincNet Frame Error Rates ($\%$) for TIMIT dataset.}
\label{tab:FER}
\begin{tabular}{cccccccccccc}
\toprule
\multirow{2}{*}{Epoch} & \multirow{2}{*}{SincNet} & \multicolumn{10}{c}{AM-SincNet} \\ \cmidrule{3-12}
 &  & m=0.35 & m=0.40 & m=0.45 & m=0.50 & m=0.55 & m=0.60 & m=0.65 & m=0.70 & m=0.75 & m=0.80 \\ \midrule
0  & \textbf{97.25} & 98.77 & 98.76 & 98.71 & 99.06 & 98.08 & 99.13 & 98.14 & 97.65 & 98.21 & 98.78 \\ \midrule
16 & 55.32 & 56.70 & 57.93 & 57.29 & 58.37 & \textbf{54.09} & 56.44 & 54.69 & 57.23 & 60.98 & 55.65 \\ \midrule
32 & 50.29 & 44.20 & 46.37 & 44.57 & \textbf{43.46} & 44.23 & 45.56 & 49.98 & 44.84 & 44.32 & 48.68 \\ \midrule
48 & 46.67 & 41.99 & 39.88 & 45.43 & 40.54 & 40.49 & 39.17 & 41.25 & 38.87 & \textbf{37.95} & 42.45 \\ \midrule
64 & 45.40 & 41.51 & 38.05 & 42.05 & 38.02 & 38.13 & 37.45 & \textbf{36.83} & 38.86 & 37.36 & 37.34 \\ \midrule
80 & 43.49 & 36.30 & 36.37 & 36.57 & 34.89 & 36.34 & 36.99 & 34.47 & \textbf{34.11} & 34.72 & 34.51 \\ \midrule
96 & 44.83 & 34.37 & 34.11 & 33.50 & 33.68 & 36.82 & 33.41 & \textbf{33.07} & 33.13 & 34.00 & 34.14 \\ \midrule
... & ... & ... & ... & ... & ... & ... & ... & ... & ... & ... & ... \\ \midrule
320 & 46.39 & 28.76 & 28.21 & 27.82 & \textbf{27.37} & 28.82 & 27.40 & 27.54 & 27.90 & 29.39 & 28.32 \\ \midrule
336 & 47.93 & 27.92 & 28.73 & 29.00 & 27.42 & 27.50 & \textbf{27.18} & 27.54 & 30.00 & 27.60 & 28.68 \\ \midrule
352 & 44.64 & 29.22 & 27.57 & 27.07 & 27.86 & 27.81 & 28.28 & 27.92 & 29.76 & \textbf{26.95} & 30.85 \\
\bottomrule
\end{tabular}
\end{table*}

\section{Results}
\label{results}

Several experiments were made to evaluate the proposed method against the traditional SincNet approach. In every one of them, the proposed AM-SincNet has shown higher accurate results. The proposed AM-SincNet method requires two more parameters, the scaling parameter $s$ and the margin parameter $m$. We have decided to use $s\!=\!30$, and we have done experiments to evaluate the influence of the margin parameter $m$ on the Frame Error Rate. 

The Table \ref{tab:FER} shows the Frame Error Rate (FER) in percentage for the original SincNet and our proposed method over 352 epochs on the test data. To verify the influence of the margin parameter on the proposed method, we performed several experiments using different values of $m$ in the range $0.35\!\leq\!m\!\leq\!0.80$. The table shows the results from the first 96 and the last 32 epochs in steps of 16. The best result from each epoch is highlighted in bold.

It is possible to see that traditional SincNet only gets better results than the proposed AM-SincNet on the first epochs when none of them have given proper training time yet. After that, on epoch 48, the original SincNet starts to converge with an FER around $46\%$, while the proposed method keeps decreasing its error throughout training.

\begin{figure}
\centering
\begin{tikzpicture}
\begin{axis}[
axis lines = left,
xlabel = Epochs,
ylabel = {FER ($\%$)},
]
\addplot [
    color=red,
    line width=0.35mm,
]coordinates {
(0, 97.25)
(8, 68.20)
(16, 55.32)
(24, 52.84)
(32, 50.29)
(40, 49.58)
(48, 46.67)
(56, 45.20)
(64, 45.40)
(72, 50.08)
(80, 43.49)
(88, 44.62)
(96, 44.83)
(104, 46.93)
(112, 47.10)
(120, 43.88)
(128, 45.71)
(136, 46.52)
(144, 44.89)
(152, 43.70)
(160, 46.54)
(168, 42.53)
(176, 43.56)
(184, 43.56)
(192, 46.83)
(200, 43.69)
(208, 43.44)
(216, 43.41)
(224, 42.55)
(232, 42.81)
(240, 44.19)
(248, 44.27)
(256, 45.32)
(264, 45.22)
(272, 43.64)
(280, 42.69)
(288, 43.64)
(296, 43.84)
(304, 44.37)
(312, 42.07)
(320, 46.39)
(328, 43.70)
(336, 47.96)
(344, 44.99)
(352, 44.64)
};
\addlegendentry{SincNet}
\addplot [
color=blue,
dashed,
line width=0.35mm,
]
coordinates {
(0, 99.06)
(8, 68.03)
(16, 58.37)
(24, 50.30)
(32, 43.46)
(40, 50.24)
(48, 40.57)
(56, 40.16)
(64, 38.02)
(72, 37.50)
(80, 34.89)
(88, 33.80)
(96, 33.68)
(104, 37.59)
(112, 32.42)
(120, 31.65)
(128, 30.49)
(136, 31.84)
(144, 30.92)
(152, 29.85)
(160, 30.65)
(168, 30.74)
(176, 30.11)
(184, 31.08)
(192, 30.73)
(200, 29.22)
(208, 29.20)
(216, 28.92)
(224, 28.16)
(232, 28.41)
(240, 28.62)
(248, 28.00)
(256, 28.64)
(264, 29.78)
(272, 28.32)
(280, 28.02)
(288, 28.41)
(296, 27.66)
(304, 28.54)
(312, 27.43)
(320, 27.37)
(328, 27.20)
(336, 27.42)
(344, 28.32)
(352, 27.86)
};
\addlegendentry{AM-SincNet (m=0.50)}
\end{axis}
\end{tikzpicture}
\caption{Comparison of Frame Error Rate ($\%$) from SincNet and AM-SincNet (m=0.50) over the training epochs for TIMIT dataset.}
\label{fig:SincvsAM-Sincm050}
\end{figure}
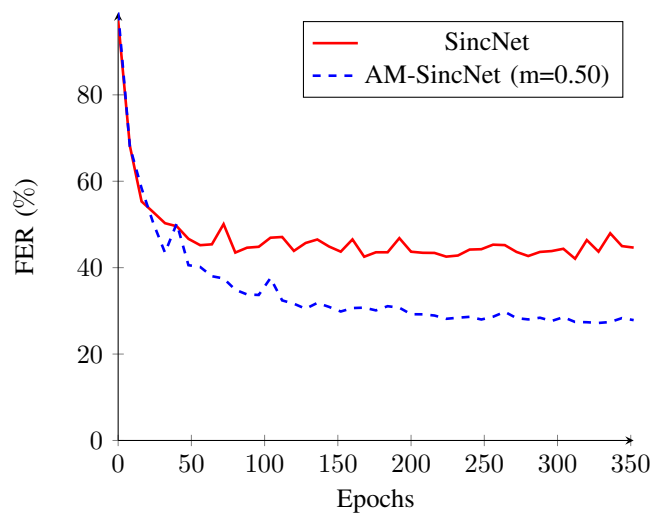

In the epoch 96, the proposed method has already an FER more than $26\%$ better than the original SincNet for almost every value of $m$ excluding $m\!=\!0.55$. The difference keeps increasing over the epochs, and at epoch 352 the proposed method has an FER of $26.95\%$ ($m\!=\!0.75$) against $44.64\%$ from SincNet, which means that at this epoch AM-SincNet has a Frame Error Rate approximately
 $40\%$ better than traditional SincNet. The Figure \ref{fig:SincvsAM-Sincm050} plots the Frame Error Rate on the test data for both methods along the training epochs. For the AM-SincNet, we used the margin parameter $m\!=\!0.50$.

From Table \ref{tab:FER}, we can also see the impact of the margin parameter $m$ on our proposed method. It is possible to see that the FER calculated for $m\!=\!0.50$ got the lowest (best) value at the epochs 32 and 320. In the same way, $m\!=\!0.55$ and $m\!=\!0.60$ got the lowest values at epochs 16 and 336, respectively. The value $m\!=\!0.65$ scores the lowest result for epochs 64 and 96, while $m\!=\!0.70$ got the lowest score at epoch 80, and $m\!=\!0.75$ reached the lowest value of epochs 48 and 352.

The $m\!=\!0.35$, $m\!=\!0.40$, $m\!=\!0.45$, and $m\!=\!0.80$ does not reach the lowest values of any epoch in this table. Although the results in Table \ref{tab:FER} may indicate that there is a golden value of $m$ which brings the best Frame Error Rate for the experiments, in fact, the difference of the FER calculated among the epochs may not be so significant. Indeed, at the end of training, all of the experiments with the AM-SincNet seem to approximate the FER to a value around $27\%$. In any case, AM-SincNet overcomes the baseline approach.

\section{Conclusion} 
\label{conclusion}

This paper has proposed a new approach for directly processing waveform audio that is highly inspirited in the neural network architecture SincNet and the Additive Margin Softmax loss function. The proposed method, AM-SincNet, has shown a Frame Error Rate about 40\% smaller than the traditional SincNet. It shows that the loss function we use on a model can have a significant impact on the expected result.

From Figure \ref{fig:SincvsAM-Sincm050}, it is possible to notice that the FER ($\%$) from the proposed method may not have converged yet on the last epochs. Thus, if the training had last more, we may have noticed an even more significant difference between both methods. The proposed method comes with two more parameters for setting when compared with the traditional SincNet, although the experiments made here show that these extra parameters can be fixed values without compromising the performance of the model.

For future work, we would like to test our method using different datasets such as VoxCeleb2 \cite{deep1}, which has over a million samples from over 6k speakers. If we increase the amount of data, the model may show a more significant result. We also intend to use more metrics such as the Classification Error Rate ($\%$) (CER) and the Equal Error Rate ($\%$) (EER) to compare the models.


\section*{Acknowledgment}

This work was supported in part by CNPq and CETENE (Brazilian research agencies). We gratefully acknowledge the support of NVIDIA Corporation with the donation of the Titan XP GPU used for this research. 

\bibliographystyle{bibtex/bst/IEEEtran}
\bibliography{bibtex/bib/IEEEexample}

\begin{thebibliography}{10}
\providecommand{\url}[1]{#1}
\csname url@samestyle\endcsname
\providecommand{\newblock}{\relax}
\providecommand{\bibinfo}[2]{#2}
\providecommand{\BIBentrySTDinterwordspacing}{\spaceskip=0pt\relax}
\providecommand{\BIBentryALTinterwordstretchfactor}{4}
\providecommand{\BIBentryALTinterwordspacing}{\spaceskip=\fontdimen2\font plus
\BIBentryALTinterwordstretchfactor\fontdimen3\font minus
  \fontdimen4\font\relax}
\providecommand{\BIBforeignlanguage}[2]{{%
\expandafter\ifx\csname l@#1\endcsname\relax
\typeout{** WARNING: IEEEtran.bst: No hyphenation pattern has been}%
\typeout{** loaded for the language `#1'. Using the pattern for}%
\typeout{** the default language instead.}%
\else
\language=\csname l@#1\endcsname
\fi
#2}}
\providecommand{\BIBdecl}{\relax}
\BIBdecl

\bibitem{Beigi}
H.~Beigi, \emph{Fundamentals of Speaker Recognition}.\hskip 1em plus 0.5em
  minus 0.4em\relax Springer Publishing Company, Incorporated, 2011.

\bibitem{Dehak10frontend}
N.~Dehak, P.~J. Kenny, R.~Dehak, P.~Dumouchel, and P.~Ouellet, ``Front end
  factor analysis for speaker verification,'' \emph{IEEE Transactions on Audio,
  Speech and Language Processing}, 2010.

\bibitem{Prince2007ProbabilisticLD}
S.~Prince and J.~H. Elder, ``Probabilistic linear discriminant analysis for
  inferences about identity,'' \emph{2007 IEEE 11th International Conference on
  Computer Vision}, pp. 1--8, 2007.

\bibitem{htPLDA}
P.~Matejka, O.~Glembek, F.~Castaldo, M.~J. Alam, O.~Plchot, P.~Kenny,
  L.~Burget, and J.~Cernocký, ``Full-covariance ubm and heavy-tailed plda in
  i-vector speaker verification,'' 05 2011, pp. 4828--4831.

\bibitem{GaussPLDA}
\BIBentryALTinterwordspacing
S.~Cumani, O.~Plchot, and P.~Laface, ``Probabilistic linear discriminant
  analysis of i-vector posterior distributions.'' in \emph{ICASSP}.\hskip 1em
  plus 0.5em minus 0.4em\relax IEEE, 2013, pp. 7644--7648. [Online]. Available:
  \url{http://dblp.uni-trier.de/db/conf/icassp/icassp2013.html#CumaniPL13}
\BIBentrySTDinterwordspacing

\bibitem{oldmethods}
W.~Campbell, D.~Sturim, D.~Reynolds, and A.~Solomonoff, ``Svm based speaker
  verification using a gmm supervector kernel and nap variability
  compensation,'' vol.~1, 06 2006, pp. I -- I.

\bibitem{oldmethods2}
P.~Kenny, G.~Boulianne, P.~Ouellet, and P.~Dumouchel, ``Joint factor analysis
  versus eigenchannels in speaker recognition,'' \emph{Audio, Speech, and
  Language Processing, IEEE Transactions on}, vol.~15, pp. 1435 -- 1447, 06
  2007.

\bibitem{oldmethods5}
S.~Cumani, O.~Plchot, and P.~Laface, ``Probabilistic linear discriminant
  analysis of i-vector posterior distributions,'' in \emph{2013 IEEE
  International Conference on Acoustics, Speech and Signal Processing}, May
  2013, pp. 7644--7648.

\bibitem{oldmethods6}
\BIBentryALTinterwordspacing
G.~Heigold, I.~Moreno, S.~Bengio, and N.~Shazeer, ``End-to-end text-dependent
  speaker verification,'' \emph{2016 IEEE International Conference on
  Acoustics, Speech and Signal Processing (ICASSP)}, Mar 2016. [Online].
  Available: \url{http://dx.doi.org/10.1109/ICASSP.2016.7472652}
\BIBentrySTDinterwordspacing

\bibitem{oldmethods7}
D.~Snyder, P.~Ghahremani, D.~Povey, D.~Garcia-Romero, Y.~Carmiel, and
  S.~Khudanpur, ``Deep neural network-based speaker embeddings for end-to-end
  speaker verification,'' in \emph{2016 IEEE Spoken Language Technology
  Workshop (SLT)}, Dec 2016, pp. 165--170.

\bibitem{advancesDeep}
M.~McLaren, Y.~Lei, and L.~Ferrer, ``Advances in deep neural network approaches
  to speaker recognition,'' in \emph{2015 IEEE International Conference on
  Acoustics, Speech and Signal Processing (ICASSP)}, April 2015, pp.
  4814--4818.

\bibitem{advancesDeep2}
F.~Richardson, D.~Reynolds, and N.~Dehak, ``Deep neural network approaches to
  speaker and language recognition,'' \emph{IEEE Signal Processing Letters},
  vol.~22, pp. 1--1, 10 2015.

\bibitem{end-to-end}
D.~Snyder, P.~Ghahremani, D.~Povey, D.~Garcia-Romero, Y.~Carmiel, and
  S.~Khudanpur, ``Deep neural network-based speaker embeddings for end-to-end
  speaker verification,'' 12 2016, pp. 165--170.

\bibitem{end-to-end2}
G.~Trigeorgis, F.~Ringeval, R.~Brueckner, E.~Marchi, M.~A. Nicolaou,
  B.~Schuller, and S.~Zafeiriou, ``Adieu features? end-to-end speech emotion
  recognition using a deep convolutional recurrent network,'' in \emph{2016
  IEEE International Conference on Acoustics, Speech and Signal Processing
  (ICASSP)}, March 2016, pp. 5200--5204.

\bibitem{end-to-end3}
J.-W. Jung, H.-S. Heo, I.-H. Yang, H.-j. Shim, and H.-J. Yu, ``A complete
  end-to-end speaker verification system using deep neural networks: From raw
  signals to verification result,'' 04 2018.

\bibitem{AM-Softmax}
F.~Wang, J.~Cheng, W.~Liu, and H.~Liu, ``Additive margin softmax for face
  verification,'' \emph{IEEE Signal Processing Letters}, vol.~25, no.~7, pp.
  926--930, July 2018.

\bibitem{SincNet}
M.~Ravanelli and Y.~Bengio, ``Speaker recognition from raw waveform with
  sincnet,'' 08 2018.

\bibitem{timit}
J.~S. Garofolo, L.~F. Lamel, W.~M. Fisher, J.~G. Fiscus, D.~S. Pallett, and
  N.~L. Dahlgren, ``Darpa timit acoustic phonetic continuous speech corpus
  cdrom,'' 1993.

\bibitem{TraditionalMethodsReview}
J.~H.~L. Hansen and T.~Hasan, ``Speaker recognition by machines and humans: A
  tutorial review,'' \emph{IEEE Signal Processing Magazine}, vol.~32, no.~6,
  pp. 74--99, Nov 2015.

\bibitem{Goodfellow-et-al-2016}
I.~Goodfellow, Y.~Bengio, and A.~Courville, \emph{Deep Learning}.\hskip 1em
  plus 0.5em minus 0.4em\relax MIT Press, 2016,
  \url{http://www.deeplearningbook.org}.

\bibitem{RavanelliBOB17}
M.~Ravanelli, P.~Brakel, M.~Omologo, and Y.~Bengio, ``A network of deep neural
  networks for distant speech recognition,'' \emph{CoRR}, vol. abs/1703.08002,
  2017.

\bibitem{deep1}
\BIBentryALTinterwordspacing
J.~S. Chung, A.~Nagrani, and A.~Zisserman, ``Voxceleb2: Deep speaker
  recognition,'' \emph{Interspeech 2018}, Sep 2018. [Online]. Available:
  \url{http://dx.doi.org/10.21437/Interspeech.2018-1929}
\BIBentrySTDinterwordspacing

\bibitem{deeo2}
C.~Li, X.~Ma, B.~Jiang, X.~Li, X.~Zhang, X.~Liu, Y.~Cao, A.~Kannan, and Z.~Zhu,
  ``Deep speaker: an end-to-end neural speaker embedding system,'' 2017.

\bibitem{hybird1}
E.~Variani, X.~Lei, E.~Mcdermott, I.~L. Moreno, and J.~Gonzalez-dominguez,
  ``Deep neural networks for small footprint text-dependent speaker
  verification.''

\bibitem{hybrid2}
Y.~Lei, N.~Scheffer, L.~Ferrer, and M.~McLaren, ``A novel scheme for speaker
  recognition using a phonetically-aware deep neural network,'' 05 2014, pp.
  1695--1699.

\bibitem{hybrid3}
S.~Hamidi~Ghalehjegh and R.~Rose, ``Deep bottleneck features for i-vector based
  text-independent speaker verification,'' 12 2015, pp. 555--560.

\bibitem{hybrid4}
\BIBentryALTinterwordspacing
D.~Snyder, D.~Garcia-Romero, D.~Povey, and S.~Khudanpur, ``Deep neural network
  embeddings for text-independent speaker verification.'' in
  \emph{INTERSPEECH}, F.~Lacerda, Ed.\hskip 1em plus 0.5em minus 0.4em\relax
  ISCA, 2017, pp. 999--1003. [Online]. Available:
  \url{http://dblp.uni-trier.de/db/conf/interspeech/interspeech2017.html#SnyderGPK17}
\BIBentrySTDinterwordspacing

\bibitem{hybrid5}
D.~Snyder, D.~Garcia-Romero, G.~Sell, D.~Povey, and S.~Khudanpur, ``X-vectors:
  Robust dnn embeddings for speaker recognition,'' \emph{2018 IEEE
  International Conference on Acoustics, Speech and Signal Processing
  (ICASSP)}, pp. 5329--5333, 2018.

\bibitem{triplet}
C.~Zhang, K.~Koishida, and J.~H.~L. Hansen, ``Text-independent speaker
  verification based on triplet convolutional neural network embeddings,''
  \emph{IEEE/ACM Transactions on Audio, Speech, and Language Processing},
  vol.~26, no.~9, pp. 1633--1644, Sep. 2018.

\bibitem{leaky-ReLU}
A.~L. Maas, A.~Y. Hannun, and A.~Y. Ng, ``Rectifier nonlinearities improve
  neural network acoustic models,'' in \emph{in ICML Workshop on Deep Learning
  for Audio, Speech and Language Processing}, 2013.

\end{thebibliography}

\end{document}